\documentclass[conference]{IEEEtran}
\IEEEoverridecommandlockouts
\usepackage{amsmath,amssymb,amsfonts}
\usepackage{algorithmic}
\usepackage{graphicx}

\ifCLASSOPTIONcompsoc \usepackage[caption=false,font=normalsize,labelfont=sf,textfont=sf]{subfig}
\else
\usepackage[caption=false,font=footnotesize]{subfig}
\fi

\usepackage{accents}
\usepackage{textcomp}
\def\BibTeX{{\rm B\kern-.05em{\sc i\kern-.025em b}\kern-.08em
    T\kern-.1667em\lower.7ex\hbox{E}\kern-.125emX}}

\usepackage[notheorems,IEEEtran]{research17}
\usepackage{balance}
\usepackage{booktabs} 
\usepackage{enumitem}
\usepackage{mathtools}
\usepackage[lined,boxed,commentsnumbered,linesnumbered,ruled]{algorithm2e}

\usepackage[capitalize]{cleveref}
\crefname{equation}{\unskip}{\unskip}
\crefname{claim}{Claim}{Claims} 
\usepackage{array}
\usepackage{multirow}
\newcolumntype{C}[1]{>{\centering\arraybackslash}p{#1}}
\allowdisplaybreaks

\renewcommand{\E}[1]{\mathbb{E}\left[ #1 \right]}

\renewcommand{\Var}[1]{\operatorname{Var} \left[ #1 \right]}

\newcommand{\rmean}{\bar{r}_1(\tau)}
\newcommand{\rprocess}{r_1(\tau)}

\newcommand{\estar}{\epsilon^*}
\newcommand{\sstart}{\alpha}
\newcommand{\ssend}{\beta}
\newcommand{\ssrange}{\left[\sstart, \ssend\right]}

\renewcommand{\e}{\epsilon}

\newcommand{\m}{\mu_0}
\newcommand{\defn}{\triangleq}
\newcommand{\der}{\mathrm{d}}
\newcommand{\ti}{\to\infty}
\newcommand{\grows}{\ti}

\newcommand{\FER}{P_\mathsf{f}}
\newcommand{\BER}{P_\mathsf{b}}

\newcommand{\doublehat}[1]{\tilde{#1}}

\newcommand{\gammasingle}{\breve{\gamma}}
\newcommand{\gammadouble}{\doublehat{\gamma}}

\newcommand{\sstartdouble}{\doublehat{\sstart}}

\newcommand{\ssenddouble}{\doublehat{\ssend}}
\newcommand{\thetasingle}{\breve{\theta}}
\newcommand{\thetadouble}{\doublehat{\theta}}
\newcommand{\nusingle}{\breve{\nu}}
\newcommand{\nudouble}{\doublehat{\nu}}

\newcommand{\dv}{d_\mathsf{v}}
\newcommand{\dc}{d_\mathsf{c}}

\newcommand{\fsingle}{f^{\mathsf{(1)}}_{\tau_0}}
\newcommand{\fdouble}{f^{\mathsf{(2)}}_{\tau_0}}

\newcommand{\sstartlb}{\sstartdouble_\mathsf{LB}}

\begin{document}

\title{A Refined Scaling Law for Spatially Coupled LDPC Codes Over the Binary Erasure Channel
\thanks{This work was funded by the Swedish Research Council (grant 2016-4026).}}

\author{\IEEEauthorblockN{
    Roman Sokolovskii,
    Fredrik Brännström,
    and Alexandre Graell i Amat}
  \IEEEauthorblockA{Department of Electrical Engineering, Chalmers University of Technology,
    SE--41296 Gothenburg, Sweden}}

\maketitle

\begin{abstract}
    We propose a refined scaling law to predict the finite-length performance in the waterfall
    region of spatially coupled low-density parity-check codes over the binary erasure channel. In
    particular, we introduce some improvements to the scaling law proposed by Olmos and
    Urbanke that result in a better agreement between the predicted and simulated frame
    error rate. We also show how the scaling law can be extended to predict the bit error rate
    performance.
\end{abstract}

\section{Introduction}

Spatially coupled low-density parity-check (SC-LDPC)
\mbox{codes~\cite{Jime99,Lent10}} have received a great deal of attention in the
last years due to the remarkable threshold saturation effect, i.e., the belief propagation (BP)
decoder can achieve the maximum-a-posteriori~(MAP) threshold of the underlying uncoupled ensemble.
Threshold saturation was first shown in \cite{Lent10} and proved for the binary erasure channel
(BEC) in~\cite{Kude11}. It was later proved for the general class of binary-input
memoryless symmetric channels~\cite{Kude13}.
Another desirable property of spatial coupling is that it preserves the distance growth properties
of the underlying ensemble, i.e., an \mbox{SC-LDPC} ensemble constructed from a regular ensemble
preserves the linear growth of the minimum distance with the block length.
Spatial coupling is a very general concept, and is not limited to low-density parity-check (LDPC) codes. It has been used, e.g., in the context of
turbo~\cite{Molo17} and polar~\cite{Wang18} codes. Staircase codes~\cite{Smit12} can also
be viewed as a class of spatially coupled codes.

The BP threshold has become a parameter of substantial importance in modern coding
theory, as it allows fast code optimization (e.g., in terms of degree distribution for LDPC codes) via density evolution.
Furthermore, a code with better threshold typically also achieves  better performance in the
waterfall region for finite length.
Predicting the probability of error for a fixed code length, however, is also of great practical
interest.
A finite-length scaling for LDPC code
ensembles over the BEC, which yields accurate predictions of the frame error rate (FER)  in the
waterfall region, was proposed in \cite{Amra09}. The scaling law is based on the analysis
of the statistical evolution of the residual graph of the peeling decoder as a function of time.
Following this approach, a scaling law was proposed in~\cite{Olmo15} to predict the FER
performance of terminated \mbox{SC-LDPC} code ensembles.
It is based on modeling the stochastic process associated with the fraction of degree-one check
nodes (CNs) during the peeling decoding by an appropriately chosen Ornstein-Uhlenbeck process. An
approximation of the distribution of the first hit time of this process, i.e., of the earliest time
when the fraction of degree-one CNs reaches zero, is used to predict the probability of decoding error.
The authors derived a system of coupled differential equations, dubbed \textit{mean and covariance
evolution} \cite{Amra09}, to estimate the parameters of the Ornstein-Uhlenbeck process.
In contrast to the case of block LDPC codes, the results in~\cite{Olmo15} show a mismatch
between the predicted performance and the simulation results, which  was attributed
in~\cite{Olmo15} to the approximation of the distribution of the first hit time  of the
Ornstein-Uhlenbeck process by an exponential distribution.

In this paper,
we propose a refined scaling law for terminated SC-LDPC codes over the BEC. In particular, we model
the decoding process as two independent Ornstein-Uhlenbeck processes, in correspondence to the two
decoding waves that propagate toward the center of the coupled chain for terminated SC-LDPC codes.
This is in contrast to the scaling law in~\cite{Olmo15}, which assumes a single process.
Accordingly, the probability density function (PDF) of the first hit time can be modeled as the
convolution of two exponential PDFs, leading to the PDF of an Erlang distribution, which we use to
predict the probability of error.
We show that this refined model results in a better agreement with simulations compared to the
original scaling law in~\cite{Olmo15}.
We further improve the match between the predicted performance and simulation results by introducing
a dependency on the channel parameter of the scaling constants that can be computed from the mean
evolution.
Finally, we show that the scaling law can be easily adapted to predict the bit error rate (BER).

\section{Preliminaries}
\label{sec:preliminaries}

We consider the $(\dv,\dc,L,N)$ SC-LDPC code ensemble introduced in~\cite{Olmo15}, composed of a
sequence of $L$ $(\dv,\dc)$-regular LDPC codes, each one with $N$ variable nodes (VNs) of degree
$\dv$ and $M\!=\!\frac{\dv}{\dc}N$ CNs of degree $\dc$, where we assume $M$ is an integer.
The sequence is coupled according to the following rule: every VN at position $i$ is connected to
one randomly chosen CN at each of the positions in the range $[i,\ldots,i+\dv-1]$.
The \textit{terminated} ensemble is obtained from such a chain by appending additional $\dv-1$
positions containing CNs only. 
Note that while this coupling rule prescribes VNs to be connected to each of the $\dv$ different
positions in $[i,\ldots,i+\dv-1]$, it does not enforce any particular connectivity on the CNs within
this range.  Thus, a given CN can be connected to VNs at the same position.  This
``semi-structured'' ensemble was chosen in~\cite{Olmo15} instead of the conventional ensemble
with  \emph{smoothing} parameter \cite{Kude11} to simplify the analysis. Besides the terminated ensemble, we also consider the \textit{truncated} ensemble, where the coupled chain is truncated after $L$ positions.

We consider decoding using the peeling decoding algorithm, which has identical performance to that of
BP decoding over the BEC, but makes the analysis more tractable. At the initialization step of the
peeling decoding, all non-erased VNs and adjacent edges are removed from the Tanner graph.  Next, at
each iteration, one degree-one CN is randomly chosen. Since its neighbor VN is known (i.e., the code
bit can be recovered), the selected CN can be deleted, as well as the neighbor VN and all its
adjacent edges.
This results in a sequence of residual graphs indexed by the iteration number~$\ell$.
The decoding process stops when there are no degree-one CNs left, which occurs either because
the whole message has been successfully recovered or because decoding fails.

In the waterfall region, the primary contributor to decoding failures are large (linear sized with
respect to the code length) stopping sets~\cite{Amra09}.
Thus, the scaling law aims to estimate the probability of a linear sized number of bits remaining
unrecovered after the termination of the peeling decoding.

\subsection{Finite-Length Scaling of SC-LDPC Ensembles in~\cite{Olmo15}}
\label{sse:old}

The scaling law in~\cite{Olmo15} is based on analyzing the first hit time of the stochastic
process associated with the fraction of degree-one CNs in the residual graphs during peeling
decoding~\cite{Luby01},
\begin{equation}
\rprocess \defn \frac{1}{N}\sum_u R_{1,u}(\tau),
\end{equation}
where $\tau \defn \ell / N$ can be viewed as normalized time of the peeling decoding process,
and $R_{1,u}(\tau)$ is the number of degree-one CNs at position $u$ of the residual graph at
iteration~$\ell$.
It was shown in~\cite{Amra09,Olmo15} that the distribution of $\rprocess$ converges  to a Gaussian
distribution as $N\grows$.

The first hit time $\tau_0$ is defined as the earliest time of the peeling decoding for which the process
$\rprocess$ hits zero, i.e.,
\begin{equation}
\tau_0 \defn \min \{\tau : \rprocess = 0 \}.
\end{equation}
We denote the PDF of $\tau_0$ as $f_{\tau_0}$.

The expected number of degree-one CNs, ${\rmean \defn \E{\rprocess}}$, where the expectation is taken over the channel and ensemble realizations, exhibits a steady-state phase
where it remains essentially constant. We denote the range of $\tau$ corresponding to the steady
state as~$\ssrange$.
The underlying assumption of the scaling law in~\cite{Olmo15} is that the decoding failure is
only possible at the steady state, hence the FER is approximated as
\begin{equation}
    \FER \approx \int_{\sstart}^{\ssend} f_{\tau_0}(x) \der x.
\label{eq:fer}
\end{equation}

The decoding process $\rprocess$ in the steady state is characterized by the following parameters.
\begin{enumerate}
    \item Expectation constant $\gamma$. The value of $\rmean$ at the steady state is approximated by
        $$
        \rmean \approx \gamma (\estar - \e),
        $$
        where $\epsilon$ is the channel erasure probability and  $\estar$ denotes the BP decoding
        threshold of the given $(\dv,\dc,L)$ SC-LDPC ensemble, which can be computed via density
        evolution.
    \item Variance constant $\nu$. The variance of $\rprocess$ is also nearly constant at
        the steady state and is approximated as
        $$
        \Var{\rprocess} \approx \frac{\nu}{N}.
        $$
    \item Correlation decay constant $\theta$. For two time instants $\tau,\zeta \in \ssrange$, the
        covariance of the decoding process along iterations of the peeling decoding is assumed to
        be
        $$
        \E{\rprocess r_1(\zeta)} - \rmean \bar{r}_1(\zeta) \approx \frac{\nu}{N}\mathrm{e}^{-\theta
        \left| \zeta - \tau \right|}.
        $$
\end{enumerate}

\begin{figure}
    \centering
    \includegraphics{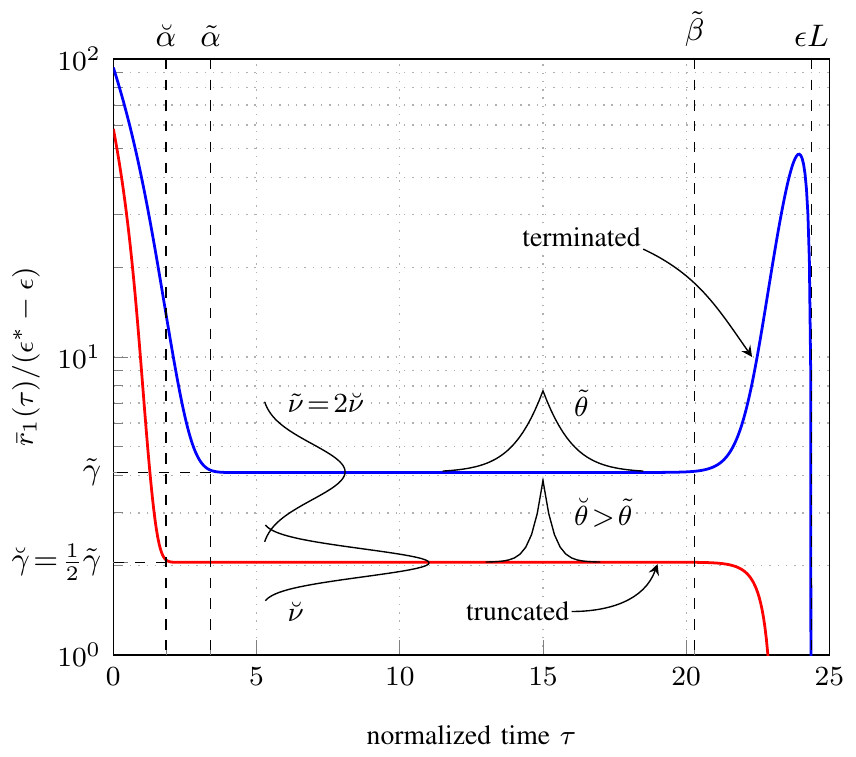}
    \vspace{-5pt}
    \caption{The evolution of the expected fraction of degree-one CNs $\rmean$ during peeling decoding,
    normalized by the distance to the BP threshold, for the
    $(5,10,L\!=\!50)$ ensemble with $\estar\!=\!0.4994$ at $\e\!=\!0.4875$.}
    \label{fig:parameters}
    \vspace{-4pt}
\end{figure}

Overall, apart from the BP threshold $\estar$, the scaling law requires the five
parameters $(\sstart,\ssend,\gamma,\nu,\theta)$. The meaning of these  parameters is illustrated in Fig.~\ref{fig:parameters} for the $(5,10,L\!=\!50)$ SC-LDPC code ensemble at
$\e\!=\!0.4875$. In the following, we denote the variables associated with the
terminated ensembles with a tilde, e.g., $\gammadouble$, and those associated with the truncated
ensembles with a breve, e.g., $\gammasingle$.

In~\cite{Olmo15}, the scaling parameters $(\gammadouble,\nudouble,\thetadouble)$ were
estimated for a particular value of $\epsilon$, namely $\e = \estar - 0.04$, by solving numerically a system of partial
differential equations called covariance evolution,
adapting the approach proposed in~\cite{Amra09} for uncoupled ensembles.
A lower bound on $\sstartdouble$, denoted as $\sstartlb$, was obtained from the density
evolution of the underlying uncoupled $(\dv,\dc)$ LDPC code ensemble by calculating the fraction of
recovered bits at the threshold~$\estar$.
The end of the steady state was approximated by $\ssenddouble = \e L$.

The decoding process $\rprocess$ at the steady state is modeled in~\cite{Olmo15} by an Ornstein-Uhlenbeck process.
Consequently, the distribution of $\tau_0$ in the steady state is approximated by the distribution
of the first hit time of the corresponding Ornstein-Uhlenbeck process, which converges 
to an exponential distribution with mean $\m$ as $N \grows$,
\begin{equation}
    \m = \frac{\sqrt{2\pi}}{\theta}\int_0^{\gamma\sqrt{N/\nu}\left( \estar - \e \right)}
    \Phi(z)\mathrm{e}^{\frac{1}{2}z^2}\der z,
    \label{eq:olmos_mean}
\end{equation}
where $\Phi(z)$ is the cumulative distribution function (CDF) of the Gaussian distribution.
Thus, the PDF of $\tau_0$ in the steady state is approximated by an exponential PDF with scale
parameter $\m$, shifted by $\sstart$ to account for the initial transient period,
\begin{equation}
    f_{\tau_0}(x)\approx\fsingle(x) \defn \m^{-1} \exp\left(-\frac{x - \sstart}{\m}\right) H(x-\sstart),
    \label{eq:olmos_pdf_approx}
\end{equation}
where $H(x)$ is the Heaviside step function.

The FER of a terminated $(\dv,\dc,L,N)$ SC-LDPC code ensemble is then estimated by using the
approximation~\eqref{eq:olmos_pdf_approx} in~\eqref{eq:fer}, resulting in~\cite{Olmo15}
\begin{equation}
\FER \approx 1 - \exp\left(-\frac{\ssend - \sstart}{\m}\right),
\label{eq:olmos_fer}
\end{equation}
with scaling parameters $(\sstartlb,\ssenddouble=\e L,\gammadouble,\nudouble,\thetadouble)$.

\section{Refined Scaling Law}
\label{sec:refinements}

We propose modeling the steady state of the stochastic process $\rprocess$ for terminated SC-LDPC
ensembles as the sum of two identical independent Ornstein-Uhlenbeck processes, to mimic the two
decoding waves moving toward the center of the chain that characterize these ensembles.
Each Ornstein-Uhlenbeck process is the same as the equivalent process for the truncated ensemble,
where only one decoding wave is present.
The decoding is successful if the two decoding waves meet, otherwise a decoding failure occurs.

\begin{figure}
    \centering
    \includegraphics{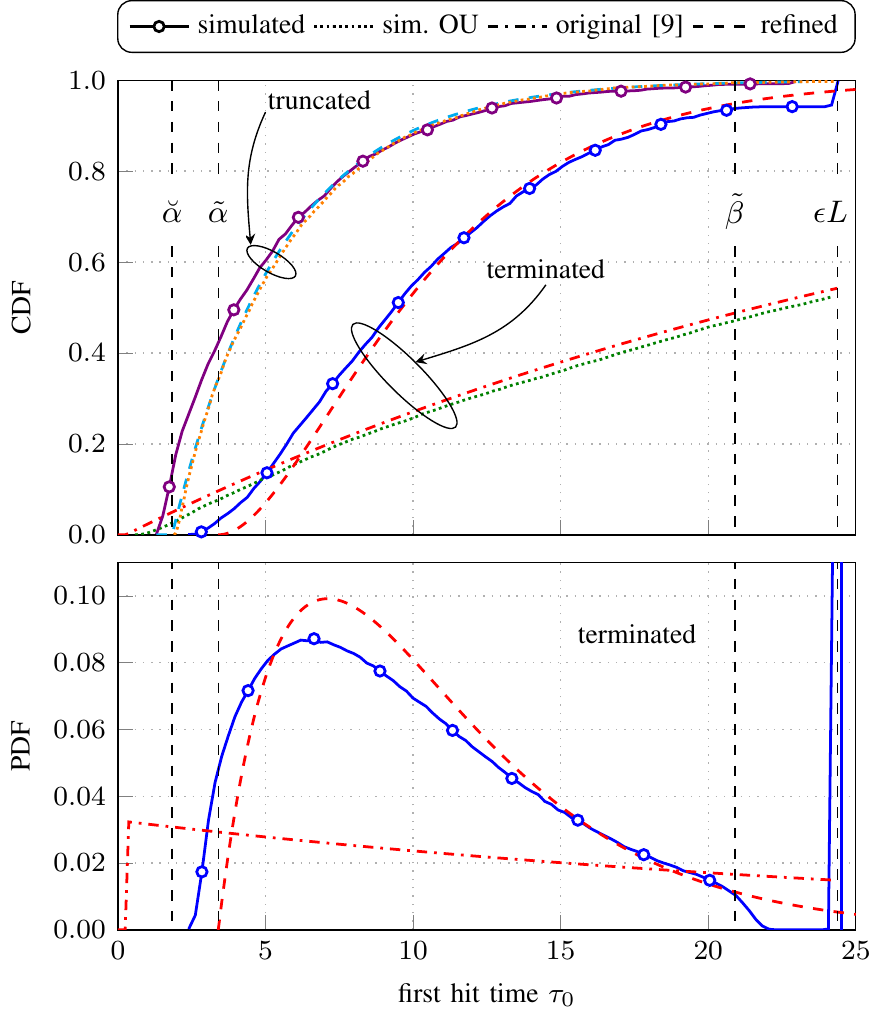}
    \vspace{-15pt}
    \caption{Comparison of the simulated and approximated CDFs of the first hit time for the terminated and
    truncated $(5,10,L\!=\!50,N\!=\!2000)$ ensembles at $\e = 0.4875$ (top). At the bottom plot,
    the corresponding PDFs for the terminated ensemble are shown.}
    \label{fig:cdf_5_10_04875}
    \vspace{-2ex}
\end{figure}

The motivation behind this model, as opposed to the model in \cite{Olmo15} based on a
single Ornstein-Uhlenbeck process, is provided in the following.
In Fig.~\ref{fig:cdf_5_10_04875} (top), we depict the simulated CDF of the first hit time 
of the peeling decoding of the terminated $(5,10,L\!=\!50,N\!=\!2000)$ SC-LDPC ensemble for
\mbox{$\e=0.4875$} (blue curve), and compare it with the simulated CDF of the first hit time of the
single Ornstein-Uhlenbeck process with appropriately chosen parameters (green dotted curve).
We also plot the CDF of the exponential distribution corresponding to the analytical approximation
$\fsingle(x)$ (see~\eqref{eq:olmos_pdf_approx}) proposed in \cite{Olmo15} (red dashdotted curve).
The corresponding PDFs are shown in Fig.~\ref{fig:cdf_5_10_04875} (bottom).
The figure shows a significant disagreement between the distributions of the first hit time of the
Ornstein-Uhlenbeck process and those of the first hit time of the peeling decoding.
Hence, a single Ornstein-Uhlenbeck process appears to be inadequate as a model for $\rprocess$ in
the steady state. In particular, it can be clearly seen in Fig.~\ref{fig:cdf_5_10_04875} (bottom) that the simulated distribution of the first hit time of the peeling decoding is not exponential. On the other hand, the agreement between the red dashdotted and green dotted curves indicates that
the exponential distribution approximates well the distribution of the first hit time of the
Ornstein-Uhlenbeck process.

In the figure, we also plot the simulated CDF of the first hit time of the peeling decoding (purple
curve), the simulated CDF of the first hit time of the Ornstein-Uhlenbeck process (orange dotted
curve) and its exponential approximation (cyan dashed curve) for the corresponding truncated
ensemble.
In this case, a much better match is observed. Hence, a single Ornstein-Uhlenbeck process models
well the behavior of the single-wave decoding process but not that of the two-wave decoding process,
motivating the proposed model based on two Ornstein-Uhlenbeck processes  for the terminated case.

\subsection{Decoding Process as Two Independent Ornstein-Uhlenbeck Processes}
\label{sse:two_waves_approx}
Since the PDF of the first hit time for the truncated ensemble (featuring a single decoding wave) is well approximated by an exponential, we model the PDF of the first hit time for the
terminated ensemble, which is characterized by two decoding waves, as the convolution of two exponential PDFs. This results in an Erlang PDF with shape parameter $2$ and scale parameter $\m$,
\begin{equation}
    \fdouble(x) \defn \m^{-2}\left(x-\sstart\right)\exp\left(-\frac{x-\sstart}{\m}\right)H(x-\sstart),
    \label{eq:terminated_pdf_approx}
\end{equation}
where $\m$ is given in~\eqref{eq:olmos_mean}. As in~\eqref{eq:olmos_pdf_approx},
the PDF is shifted by $\sstart$ so that the initial transient period before the
establishment of the steady-state regime is taken into account.

Thus, for the terminated ensemble,  we approximate the PDF of $\tau_0$ in the steady state as
\begin{equation}
\label{eq:approx2}
f_{\tau_0}(x)\approx\fdouble(x).
\end{equation}
  Using \eqref{eq:terminated_pdf_approx}--\eqref{eq:approx2} in~\eqref{eq:fer}, the FER of the terminated SC-LDPC code ensemble can then be approximated as
\begin{equation}
    \FER \approx 1 - \left( 1 + \frac{\ssend - \sstart}{\m} \right)
    \exp \left( - \frac{\ssend - \sstart}{\m} \right).
    \label{eq:our_fer}
\end{equation}

We remark that in the refined scaling law \eqref{eq:our_fer}, the scaling parameters $(\gamma,\nu,\theta)$ are those corresponding to a single
decoding wave, hence they should be estimated for the truncated ensemble.
On the other hand, the time interval $\ssrange$ of the steady state determines the times when the
two-wave regime is established and ended, thus $\sstart$ and $\ssend$ must be estimated
from the evolution of $\rmean$ for the terminated ensemble.
To summarize, the FER in~\eqref{eq:our_fer} 
should be evaluated using the scaling parameters
$(\sstartdouble,\ssenddouble,\gammasingle,\nusingle,\thetasingle)$.

It is important to emphasize that the authors of~\cite{Olmo15} mentioned that considering the
decoding process as two processes corresponding to the two decoding waves would affect the scaling
constants $\nu,\gamma$, and $\theta$, from which the constants for the combined process (still with
the first hit time modeled as an exponential distribution) could be obtained.
Here, however, we argue that the separate treatment of the two decoding waves leads not only to a
change in the scaling constants, but also to the change of the distribution of the first hit time of
the combined process from an exponential to an Erlang distribution.

\subsection{Dependence of the Scaling Parameters on the Channel Parameter}
\label{sse:additional_refinements}

The scaling law in~\cite{Olmo15} models the scaling parameters $\sstart,\gamma,\nu,$ and
$\theta$ as constants independent of the channel parameter~$\e$.
However, their estimation for different values of $\e$ yields different parameters, which indicates
that they are, in fact, dependent on $\e$.
Thus, it is preferable to calculate these parameters for each $\e$ separately.

The parameters $\sstart, \ssend,$ and $\gamma$ depend on the evolution of $\rmean$, and can be
obtained from the mean evolution in~\cite{Olmo15}.
We remark that the range of the steady state $\ssrange$ depends not only on $\e$ but also on the
length of the coupled chain $L$.
For each $(\dv,\dc,L)$ SC-LDPC ensemble, we estimate $\sstart,\ssend,$ and $\gamma$ from the
evolution of $\rmean$ for a number of channel parameters $\e$ and obtain the intermediate values by
linear interpolation.
Introducing the dependence of $\sstart$, $\ssend$, and $\gamma$ on $\epsilon$ slightly improves the
prediction of the FER.

Estimating $\nu$ and $\theta$ as a function of $\e$, on the other hand, requires numerically solving the
full covariance evolution for each $\e$, which is significantly more complex than the mean evolution
and thus renders the approach infeasible. Therefore, as in~\cite{Olmo15} we treat the variance parameter $\nu$ and the covariance decay parameter
$\theta$ as ensemble-dependent constants and use the same values of~$\nu$ and~$\theta$ for all
channel parameters~$\e$.
In this work, we resort to Monte-Carlo simulations of the peeling decoding process to estimate
the constants $\nu$ and $\theta$.
We set $N=10^4$ and the highest $\e$ for which the system operates in an effectively error-free
regime.

In Fig.~\ref{fig:cdf_5_10_04875}, we plot the approximation of the CDF and the PDF of
the first hit time for the terminated ensemble (red dashed line) by the Erlang distribution
with parameters $\sstartdouble$, $\ssenddouble$, and $\gammasingle$ computed for $\epsilon= 0.4875$.
The approximation is in good agreement with the simulated distributions of the first hit
time of the peeling decoding, which supports the proposed model.

We remark that making $\sstart$, $\ssend$, and $\gamma$ dependent on $\epsilon$ does not improve the
prediction of the original scaling law in~\cite{Olmo15}, further indicating that the main
disagreement comes from modeling the decoding process with a single Ornstein-Uhlenbeck process
instead of two processes. 

\subsection{Scaling Law to Predict the Bit Error Rate}

It is possible to apply the described framework to estimate the BER performance of an SC-LDPC code
ensemble.
Suppose the peeling decoder halted at normalized time $\tau_0\!=\!x$.
In that case it would have performed $xN$ iterations before the decoding failure, so approximately
$\e LN-xN$ out of $LN$ bits would remain unrecovered.
Accordingly, the BER can be approximated as
\begin{equation}
\label{eq:ber}
    \BER \approx \int_{\sstart}^{\ssend}  \left(\e - \frac{x}{L}\right) f_{\tau_0}(x) \der x.
\end{equation}
Using the approximation~\eqref{eq:terminated_pdf_approx}--\eqref{eq:approx2} in~\eqref{eq:ber}, we
obtain the predicted BER for the terminated ensemble as
\begin{align}
    \BER &\approx \exp \left(-\frac{\ssend - \sstart}{\m}\right) \cdot \frac{\ssend^2 + \sstart\e L - \left( \e L + \sstart - 2\m \right)
    \left( \ssend + \m \right)}{\m L} \nonumber \\
    &\mathrel{\phantom{=}} + \frac{\e L - \sstart - 2\m}{L}. \label{eq:our_ber}
\end{align}

In summary, the prediction of the FER and BER for the terminated $(\dv,\dc,L,N)$ SC-LDPC code
ensemble is given by~\eqref{eq:our_fer} and~\eqref{eq:our_ber}, respectively, with parameters
$(\sstartdouble_\e,\ssenddouble_\e,\gammasingle_\e,\nusingle,\thetasingle)$.
The dependence of the parameters on $\epsilon$ is highlighted with the subscript $\e$.

\section{Numerical Results}
\label{sec:numerical-results}

\begin{figure}
    \centering
    \includegraphics{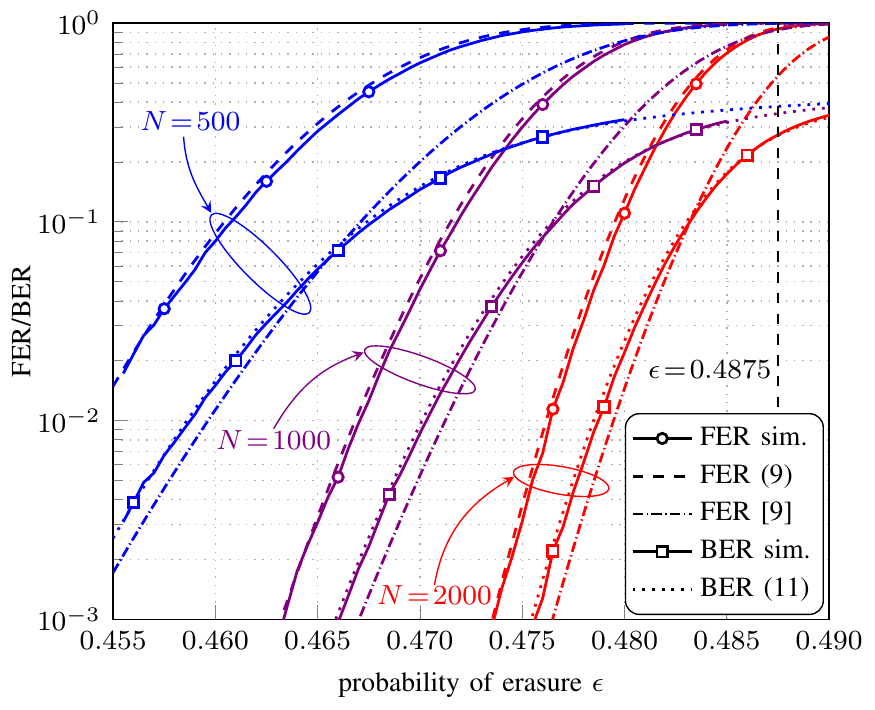}
    \vspace{-15pt}
    \caption{Simulated FER and BER curves and the corresponding analytical approximations
    for the terminated $(5,10,L\!=\!50,N)$ ensemble with \mbox{$\estar\!=\!0.4994$} for different $N$.
    The parameters $\nusingle$ and $\thetasingle$ are estimated at $\e = 0.485$ to be
    $\nusingle \approx 0.424$ and $\thetasingle \approx 1.64$.}
    \label{fig:our_approx_5_10}
    \vspace{-2ex}
\end{figure}

\begin{figure}
    \centering
    \includegraphics{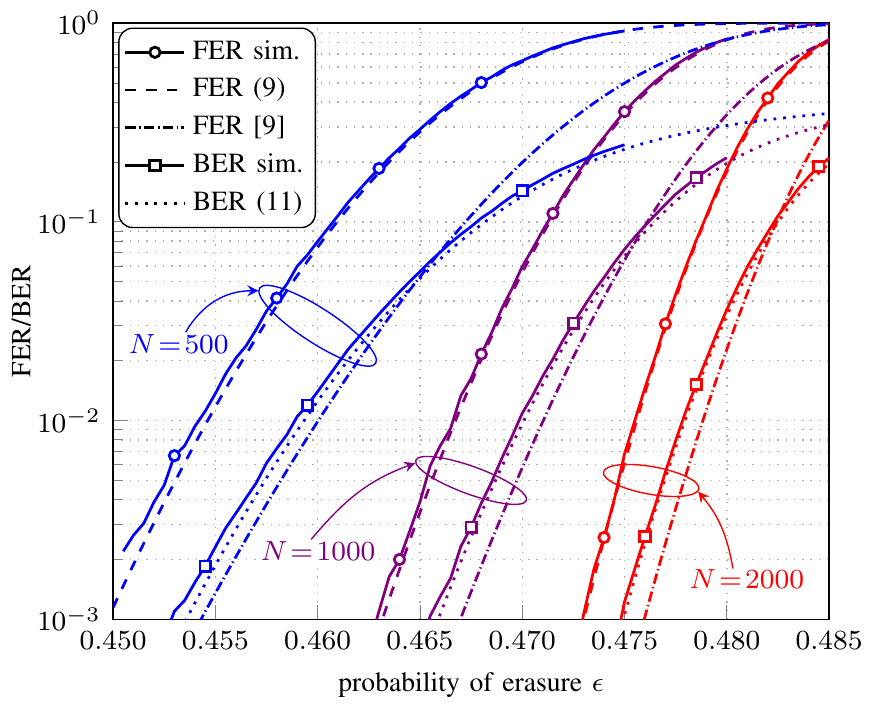}
    \vspace{-15pt}
    \caption{Simulated FER and BER curves and the corresponding analytical approximations
    for the terminated $(4,8,L\!=\!50,N)$ ensemble with \mbox{$\estar\!=\!0.4977$} for different $N$.
    The parameters $\nusingle$ and $\thetasingle$ are estimated at $\e = 0.48$ to be
    $\nusingle \approx 0.406$ and $\thetasingle \approx 1.47$.}
    \label{fig:our_approx_4_8_term}
    \vspace{-2ex}
\end{figure}

\begin{figure}
    \centering
    \includegraphics{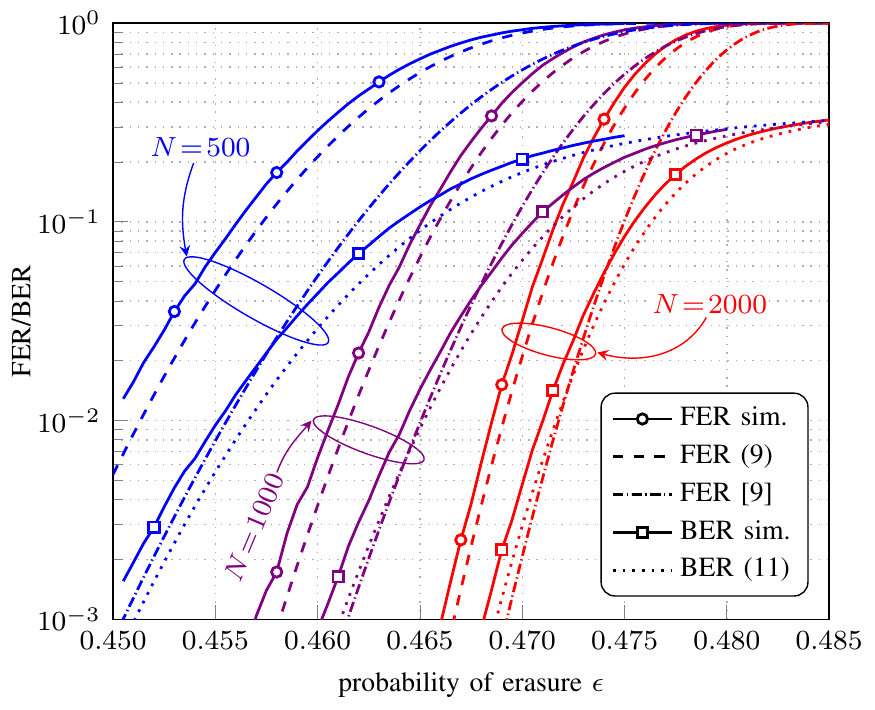}
    \vspace{-15pt}
    \caption{Simulated FER and BER curves and the corresponding analytical approximations
    for the terminated $(3,6,L\!=\!50,N)$ ensemble with \mbox{$\estar\!=\!0.4881$} for different $N$.
    The parameters $\nusingle$ and $\thetasingle$ are estimated at $\e = 0.475$ to be
    $\nusingle \approx 0.338$ and $\thetasingle \approx 1.28$.}
    \label{fig:our_approx_3_6_term}
    \vspace{-2ex}
\end{figure}

In Fig.~\ref{fig:our_approx_5_10}, we compare the simulated FER and BER
performance with the analytical approximations in~\eqref{eq:our_fer} and~\eqref{eq:our_ber} for the
terminated ${(5,10,L\!=\!50,N)}$ SC-LDPC code ensemble with $N=500$, $1000$, and $2000$.
For comparison purposes, we also include the FER predicted using the scaling law in~\cite{Olmo15} (see Section~\ref{sse:old}).
We observe that the proposed refined scaling law yields a significantly improved prediction of the FER
performance. The BER performance is also very well predicted.
We remark that, similar to the approach used in~\cite{Amra09}, to remove the effect of the error floor we consider an expurgated ensemble by
ignoring all failures involving only size-$2$ stopping sets in the calculation of the simulated
error rates.

In
Fig.~\ref{fig:our_approx_4_8_term}, we show the FER and BER performance of the terminated
${(4,8,L\!=\!50,N)}$ SC-LDPC code ensemble. Similar to the ${(5,10,L\!=\!50,N)}$ ensemble, a very good match is observed between the predicted curves and the simulation results.

Finally, in Fig.~\ref{fig:our_approx_3_6_term} we give results for the terminated $(3,6,L\!=\!50,N)$ SC-LDPC ensemble.
In this case, while the refined scaling law yields a significantly better prediction than the scaling  law in~\cite{Olmo15},  
a gap between the predicted and simulated curves remains.
In general, we observed that, irrespective of the code rate, for ensembles with VN degree $\dv\!\ge\!4$
the predicted error rates are very accurate, whereas for $\dv\!=\!3$ a gap appears, albeit
significantly smaller than that for the original scaling law in~\cite{Olmo15}.

\section{Conclusion}
\label{sec:conclusion}

We proposed a refined finite length scaling law for terminated SC-LDPC codes over the BEC.
The proposed scaling models the decoding process as two independent Ornstein-Uhlenbeck processes, corresponding to the
two decoding waves moving from the boundaries toward the center of the coupled chain.
This model accurately predicts the frame and bit error rate performance of terminated SC-LDPC
ensembles with VN degree $\dv\!\ge\!4$.
Closing the small gap in the case of $\dv\!=\!3$ is an interesting research problem, which may
be partially addressed by taking into account the possibility of a decoding failure outside of the
steady state region.


\bibliographystyle{IEEEtran}

\end{document}